\documentclass[aps,prb,twocolumn,showpacs]{revtex4}
\usepackage{bm}
\usepackage{graphicx}
\usepackage{amsmath}
\usepackage{eufrak}
\usepackage{color}
\newcommand{\nix}[1]{}

\begin{document}

\title{Magneto-gyrotropic photogalvanic effect and spin dephasing in
(110)-grown GaAs/AlGaAs quantum well structures}
\author{P.~Olbrich$^1$, J. Allerdings $^1$, V.V.~Bel'kov$^{1,2}$,
S.A.~Tarasenko$^2$, D.~Schuh$^1$, W.~Wegscheider$^1$, T.~Korn$^{1}$,   C.~Sch{\"u}ller$^{1}$,
D.~Weiss$^{1}$, and S.D.~Ganichev$^{1}$}
\affiliation{$^1$Terahertz Center, University of Regensburg,
93040 Regensburg, Germany}
\affiliation{$^2$A.F.~Ioffe Physical-Technical Institute, Russian Academy of
Sciences, 194021 St.~Petersburg, Russia}

\begin{abstract}
We report on the magneto-gyrotropic photogalvanic effect (MPGE)
in $n$-doped (110)-grown GaAs/AlGaAs quantum-well (QW) structures 
caused by free-carrier absorption of
terahertz radiation in the presence of a magnetic field.
The photocurrent behavior upon variation of the radiation polarization state,
magnetic field orientation and temperature is studied. 
The developed theory of MPGE describes well all experimental results.
It is demonstrated that the structure inversion asymmetry 
can be controllably tuned to zero by variation of 
the delta-doping layer positions. 
For the in-plane magnetic field the photocurrent is only observed 
in asymmetric structures but vanishes in symmetrically doped QWs.
Applying time-resolved Kerr rotation and
polarized luminescence we investigate the spin relaxation
in QWs for various excitation levels. Our data
confirm that in symmetrically doped QWs the spin relaxation time is
maximal, therefore, these structures set the upper limit of spin dephasing in
GaAs/AlGaAs QWs.
\end{abstract}
% 73.50.Pz Photoconduction and photovoltaic effects
% 72.25.Fe Optical creation of spin polarized carriers
% 72.25.Rb Spin relaxation and scattering
% 78.67.De Quantum wells

\maketitle

\section{Introduction}

Zincblende-structure-based  quantum wells (QWs)
grown on (110)-oriented substrates recently
attracted considerable attention. The particular feature of such structures is
their extraordinarily slow spin dephasing, being of importance for spin transport in
spintronic devices~\cite{Ohno1999,Harley03,Dohrmann,x4}. The reason for the
long spin lifetime of several nanoseconds  even at room
temperature is the suppression of the D'yakonov-Perel' mechanism of spin
relaxation in symmetrical 
(110)-grown heterostructures.~\cite{DK86}  In QWs of such
crystallographic orientation, the effective magnetic field induced by the bulk
inversion asymmetry (BIA) points along the growth axis and, therefore, does not
lead to the relaxation of spins oriented along this direction. 
However, in asymmetrical structures, where the structure
inversion asymmetry (SIA) is present,  Rashba spin-orbit
coupling~(for review see~\cite{Ch7Bychkov84p78,Dyakonov08,Fabian08}) 
induces an in-plane effective magnetic field, thus speeding-up
spin dephasing.
Experimental access to the symmetry, spin splitting of the band structure, 
etc., is provided by  the
magneto-gyrotropic photogalvanic effect (MPGE)~\cite{PRL08,BelkovGanichev}.
The MPGE stands for a photocurrent generation 
which requires simultaneously gyrotropy and the presence of
a magnetic field.~\cite{BelkovGanichev,Ivchenkobook2,GanichevPrettl} The gyrotropic point
group symmetry makes no difference between components of axial and polar
vectors, and hence allows an electric current $j_{\alpha} \propto I B_{\beta}$,
where $I$ is the light intensity inside the sample and $B_{\beta}$ are
components of the applied magnetic field.  The microscopic model
of the MPGE is based on the asymmetry of photoexcitation and/or relaxation
processes in low-dimensional systems with bulk or structure inversion
asymmetries~\cite{naturephysics06,tarasenko08}. 

In this paper, we present an experimental and theoretical  study
of the MPGE induced by Drude absorption of
terahertz (THz) radiation in GaAs/AlGaAs QWs grown on (110)-oriented substrates.
In contrast to  interband  optical
transitions, here we deal with monopolar currents
because only one type of carriers, conduction electrons, are
 involved in the photoexcitation. The paper is organized as
following. 
In Sec.~\ref{sphenomen}, 
the macroscopic features of the magneto-gyrotropic effect, e.g., the possibility to
generate a photocurrent  in various experimental geometries and
its behavior upon variation of the radiation polarization, 
are described in the frame of the phenomenological theory.
In Sections~\ref{sexperiment} and~\ref{sresults}, 
we give a short account of the experimental technique,
present the experimental results on the photocurrents
and discuss them in view of the theoretical background. Here, we also discuss
applications of the MPGE, in particular, 
for the study of BIA and SIA
responsible for the spin splitting of subbands in $\bm{k}$-space. 
Preliminary results on the study of BIA and SIA are published in Ref.~[\onlinecite{PRL08}].
In Sec.~\ref{stimeresolved}, the experimental data on spin
relaxation obtained by means of time-resolved Kerr rotation and polarized
luminescence are discussed and compared with the MPGE data.
Finally, in Sec.~\ref{scpge} we present a short account on
magnetic field independent linear and circular photogalvanic currents, which
can also be generated in (110)-grown QWs at normal incidence and should be
taken into account in studying the MPGE.

\section{Phenomenological theory}\label{sphenomen}

The phenomenological theory of magneto-gyrotropic effects describes
dependencies of the photocurrent magnitude and direction 
on the radiation polarization state and the orientation of the magnetic field
with respect to the crystallographic axes. This 
theory operates with measurable physical quantities, 
such as electric current, magnetic field, and
light polarization, and does not require a knowledge of the microscopic
mechanisms involved.

Within linear approximation in the magnetic field strength $\bm{B}$, the MPGE
at normal incidence is described  by~\cite{Belkov05}
\begin{equation} \label{phen0}
j^{{\rm \: MPGE}}_\alpha = \sum_{\beta\gamma\delta}
\phi_{\alpha\beta\gamma\delta}\:B_\beta\:I\:\frac{\left(e_\gamma e^*_\delta +
e_\delta  e^*_\gamma\right)}{2}
\end{equation}
\[
+ \sum_{\beta\gamma} \mu_{\alpha\beta\gamma}\, B_{\beta} \, \hat{e}_{\gamma} \,
I P_{circ}\:,
\]
where $\bm{\phi}$ is a fourth rank pseudo-tensor being symmetric in the last
two indices, $\bm{e}=\bf{E} / \left| \bf {E} \right|$ is the (complex) unit vector of the light polarization, $\bf E$ is the radiation electric field,
$\hat{\bm{e}}$ is the unit vector pointing in the light propagation direction
and $P_{circ}$ is the radiation circular polarization degree.
The pseudo-tensor $\bm{\phi}$ describes photocurrents which can be induced
by linearly polarized or unpolarized radiation while $\mu_{\alpha\beta\gamma}$
stands for the light helicity dependent photocurrent which reverses its
direction upon switching the sign of the circular polarization.

%%%%%%%%%%%%%%%%%%%FIGURE_01%%%%%%%%%%%%%%%%%%%%%%%%%%%%%%%%%%%%%%%%%%%%%%%%%%%%%
\begin{figure}[t]
\includegraphics[width=0.95\linewidth]{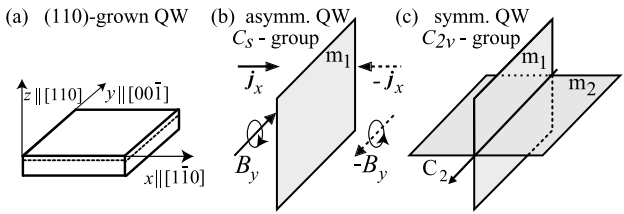}
\caption{(a) Coordinate system together with sample sketch, (b) and (c) mirror
planes in asymmetrical and symmetrical QW grown along $z ||$ [110],
respectively. Arrows in the drawing (b) show that the reflection in the
plane $m_1$ changes the sign of both polar vector component $j_x$ and axial
vector component $B_y$, demonstrating that the linear coupling $j_x \propto
B_y$ is allowed in asymmetrical QWs.
In symmetrical QWs [sketch (c)], the linear coupling of the in-plane current and the in-plane
magnetic field is forbidden because the reflection in the plane $m_2$ does not
modify ${\bm j}$ but changes the sign of the in-plane components of the axial
vector ${\bm B}$.
} \label{figure01}
\end{figure}

As in the experiment described below, we consider zinc-blend structure based quantum
wells grown on (110)-oriented substrates and normal incidence of the light.
Depending on the equivalence or non-equivalence of the QW interfaces, the
structure symmetry may belong to one of the point groups:
C$_{2v}$ or C$_s$, respectively.~\cite{footnote1} The symmetry elements of
symmetrical and asymmetrical QWs are shown in Fig.~\ref{figure01}, where the
coordinate frame with the in-plane axes $x \parallel [1\bar{1}0]$ and $y
\parallel [00 \bar{1}]$, and the growth direction $z \parallel [110]$
is used for convenience. 

The phenomenological equation~(\ref{phen0}) shows that the photocurrent can
only occur for those components of the field $\bm{B}$ and polarization vector
$\bm{e}$ whose products transform as the in-plane components of $\bm j$ for all
symmetry operations. 
Let us consider it for unpolarized radiation. In this particular case, 
 the photocurrent is  determined solely by the coupling of
the polar vector $\bm{j}$ to the axial vector $\bm{B}$
because $\left(e_\gamma e^*_\delta + e_\delta  e^*_\gamma\right) =
\delta_{\gamma,\delta}$ becomes an invariant
and $P_{circ}=0$. In the asymmetric structures ($C_{s}$ point group) the only
symmetry element, apart from identity, 
is the reflection plane $m_1$ normal to the $x$ axis. The reflection in
 $m_1$ transforms the current component $j_x$ and the magnetic
field component  $B_{y}$
the same way  ($j_x \rightarrow -j_x$, $B_{y} \rightarrow -B_{y}$), see Fig.~\ref{figure01}b. Therefore, the coupling $j_x \propto
B_{y}$ is allowed.
The same arguments hold for $j_y \propto B_{x}$ and $j_x \propto B_z$. 
As a result, the generation of a magnetic field-induced photocurrent is
possible for both, in-plane and out-of-plane magnetic fields. 
Symmetric (110)-grown QWs, which belong to the
higher point-group symmetry $C_{2v}$, contain an additional mirror plane $m_2$
being parallel to the interface plane, see Fig.~\ref{figure01}c. The
reflection in $m_2$ does not modify the in-plane components of the
polar vector ${\bm j}$ but changes the polarity of the in-plane components of the
axial vector ${\bm B}$. Therefore, a
linear coupling of the current and the in-plane magnetic field is
forbidden. The coupling of $j_x$ and $B_z$, however, is allowed for reflections
in both $m_1$ and $m_2$ planes demonstrating that a
photocurrent $j_{x}$ can be induced in the presence of a magnetic field $B_z$,
even in symmetric (110)-oriented QWs. 
This analysis shows that the MPGE is an ideal tool to probe the
symmetry of (110)-grown QWs. For in-plane magnetic field, the photocurrent
can only be observed in asymmetric structures but vanishes if the QWs are
symmetric.

For polarized radiation the phenomenological theory
should also take into account components of the % vectors of radiation
polarization vector. This results in additional polarization dependent
contributions to the photocurrent. Below we present results of such an
analysis of the MPGE describing the polarization dependence of the photocurrent
for symmetrical and asymmetrical structures.

\subsection{Asymmetrical structures, C$_{{\rm s}}$ point group}

In asymmetrical structures, the MPGE induced by normally-incident linearly-polarized light is
generally described by 9 linearly independent constants $S_1 \ldots S_9$, see Table \ref{table01}:
\begin{equation}\label{MPGE_Cs}
j_{x}^{{\rm \: MPGE}}/I =  S_1 B_y + S_2 B_y (|e_x|^2-|e_y|^2)
\end{equation}
\[
+ S_3 B_x (e_x e_y^* + e_y e_x^*) + S_4 B_z + S_5 B_z (|e_x|^2-|e_y|^2) \:,
\]
\begin{equation}
j_{y}^{{\rm \: MPGE}}/I =  S_6 B_x + S_7 B_x (|e_x|^2-|e_y|^2) \nonumber
\end{equation}
\[
+ S_8 B_y (e_x e_y^* + e_y e_x^*) + S_9 B_z (e_x e_y^* + e_y e_x^*) \:.
\]
The polarization dependence of the photocurrent is given by
\begin{equation}
\label{trigon}
|e_x|^2-|e_y|^2 = \cos 2\alpha
\,\,,\;\;\; e_x e_y^* + e_y e_x^* =  \sin 2\alpha\,\,,
\end{equation}
, where $\alpha$ is the angle between the plane of linear
polarization and the $x$ axis. We note that the first terms on the right-hand
side of both Eqs.~(\ref{MPGE_Cs}) do not depend on the radiation polarization
and describe the currents generated by unpolarized light.

The magnetic field-induced photocurrent can also be excited by
elliptically or circularly polarized radiation.
In this case, Eqs.~(\ref{MPGE_Cs}) remain valid and describe the current
independent of the sign of circular polarization.  The radiation helicity,
however, gives rise to additional current contributions given by
\begin{eqnarray}\label{MPGE_Cs2}
j_{x}^{{\rm \: MPGE, \: circ}}/I &=&  S_{10} B_{x} P_{{\rm circ}}  \:, \nonumber \\
j_{y}^{{\rm \: MPGE, \: circ}}/I &=& S_{11} B_y  P_{{\rm circ}} + S_{12} B_z
P_{{\rm circ}} \:.
\end{eqnarray}
In the experiments, elliptically and, in particular, circularly polarized
radiation is achieved by passing the laser radiation, initially linearly polarized, e.g.,
along the $x$ axis, through a $\lambda/4$-plate. The rotation of the plate
results in a variation of both linear and circular polarizations as follows
\begin{equation}
\label{plin} P_{\rm lin} \equiv  (e_{x} e^*_{y} + e_{y} e^*_{x}) = \frac12
\sin{4 \varphi}\:,\:
\end{equation}
\vspace{-0.7cm}
\begin{equation} \label{plinprime}
P'_{\rm lin} \equiv (|e_{x}|^2 - |e_{y}|^2) = \frac{1 + \cos{4 \varphi}
}{2}\:,\:
\end{equation}
\vspace{-0.7cm}
\begin{equation}
\label{circ}
 P_{\rm circ} = \sin{2 \varphi} \:,
\end{equation}
where $\varphi$ is the angle between the optical axis of the $\lambda/4$ plate
and the direction of the initial polarization. Two Stokes parameters
 $P_{\rm lin}$ and $P'_{\rm lin}$ describe the degrees of linear
polarization along the bisector (xy) and the $x$ axis, respectively, and vanish
if the radiation is circularly polarized. The third Stokes parameter $P_{\rm
circ}$ describes the radiation helicity. It is zero for linearly polarized
radiation and reaches $\pm 1$ for  left- or right-handed circular polarization.

%%%%%TABLE_01%%%%%%%%%%%%%%%%%%%%%%%%%%%%%%%%%%%%%%%%%%%%%%%%%%%%%%%%%%%%%
\begin{table}[t]
\renewcommand{\arraystretch}{1.5}
\begin{tabular}{|r@{=}l|r@{=}l|}
\hline

$S_1$ & $\frac{1}{2}(\phi_{xyxx}+ \phi_{xyyy})$ & $S_2$ &
$\frac{1}{2}(\phi_{xyxx}- \phi_{xyyy})$ \\\hline

$S_3$ & $\phi_{xxxy}=\phi_{xxyx}$ & $S_4$ & $\frac{1}{2}(\phi_{xzxx}+
\phi_{xzyy})$ \\\hline

$S_5$ & $\frac{1}{2}(\phi_{xzxx}- \phi_{xzyy})$ & $S_6$ &
$\frac{1}{2}(\phi_{yxxx}+ \phi_{yxyy})$
\\\hline

$S_7$ & $\frac{1}{2}(\phi_{yxxx}- \phi_{yxyy})$ & $S_8$ &
$\phi_{yyxy}=\phi_{yyyx}$
\\\hline

$S_9$ & $\phi_{yzxy}=\phi_{yzyx}$ & $S_{10}$ & $\mu_{xxz}$
\\\hline

$S_{11}$ & $\mu_{yyz}$ & $S_{12}$ & $\mu_{yzz}$
\\\hline
\end{tabular}
\caption{Definition of the parameters $S_i$ ($i=1\dots9$) in Eqs.~\protect (\ref{MPGE_Cs})
and~\protect (\ref{MPGE_Cs2})
in terms of non-zero components of the tensors $\bm{\phi}$ and $\bm{\mu}$ for
asymmetric (110)-grown QWs. Normal incidence of radiation along the $z$
axis is assumed.} \label{table01}
\end{table}

\subsection{Symmetrical structures, C$_{2{\rm v}}$ point group}

As addressed above,  in (110)-grown structures with equivalent
interfaces, one of the two mirror planes lies 
in the QW plane. 
A reflection  in  this mirror plane does not modify the in-plane
components of polar vectors ($e_x$, $e_y$, $j_x$, $j_y$) and the out-of-plane
component of axial vectors ($B_z$), but changes the sign of in-plane components
of axial vectors ($B_x$, $B_y$). 
Therefore, in symmetrical (110)-oriented QW structures, the MPGE
induced by normally-incident light
is described by
Eqs.~(\ref{MPGE_Cs}) together with Eqs.~(\ref{MPGE_Cs2}), where $S_1,S_2,S_3,S_6,S_7,S_8,S_{10},S_{11}=0$, i.e, by
\begin{eqnarray}\label{MPGE_C2v}
j_{x}^{{\rm \: MPGE}}/I &=&  S_4 B_z + S_5 B_z (|e_x|^2-|e_y|^2) \:, \nonumber \\
j_{y}^{{\rm \: MPGE}}/I &=&  S_9 B_z (e_x e_y^* + e_y e_x^*) + S_{12} B_z P_{{\rm circ}} \:.
\end{eqnarray}

\subsection{Linear and circular photogalvanic effects in (110)-grown QWs}

Due to the specific crystallographic orientation of (110)-grown QWs, photogalvanic currents at normal incidence of radiation are allowed even at zero magnetic field. 
They comprise the linear and the circular photogalvanic effect. 
For normal incidence of the radiation, the photocurrents are given
by~\cite{Ivchenkobook2}
\begin{eqnarray}
\label{CPGELPGE}
j_{x}^{{\rm \: PGE}}/I &=&  C_1 (e_{x} e_{y}^* + e_{y} e_{x}^*) + C_2 P_{{\rm circ}} \:, \\
j_{y}^{{\rm \: PGE}}/I &=&  C_3  +  C_4  (|e_{x}|^2 - |e_{y}|^2) \:, \nonumber
\end{eqnarray}
where the parameters $C_1$, $C_3$, and $C_4$ describe the linear photogalvanic
effect, while $C_2$ stands for the circular photocurrent, which reverses its
direction upon switching the light helicity sign. From Eq.~(\ref{CPGELPGE})
follows that the excitation with unpolarized light in the absence of a magnetic
field can lead to an electric current along the $y$ axis only. We note that
oblique incidence gives additional roots to photogalvanic effects and may also
cause the linear and circular photon drag effect~\cite{Shalygin06}.

\section{Experimental and samples}\label{sexperiment}

Magnetic field-induced photocurrents  in our experiments were induced by
indirect $intra$-subband (Drude-like) optical transitions in the lowest
size-quantized subband. We used for optical excitation the emission from a
terahertz (THz) molecular laser, optically pumped by a TEA CO$_2$
laser~\cite{GanichevPrettl}. With  NH$_3$ as active gas, 100~ns pulses of
linearly polarized or unpolarized radiation with peak power $\sim$10~kW 
are obtained at  wavelengths $\lambda=$ 90, 148 and 280~$\mu$m 
(corresponding photon energies $\hbar \omega$ are 13.7~meV, 8.4~meV and 4.4~meV). 
The terahertz radiation induces free carrier absorption in the lowest conduction subband $e1$
because the photon energies are
smaller than the subband separation and much larger than the $\bm k$-linear
spin splitting. The samples were irradiated along the growth direction.

\begin{figure}%[width=4cm]
\includegraphics[width=0.8\linewidth]{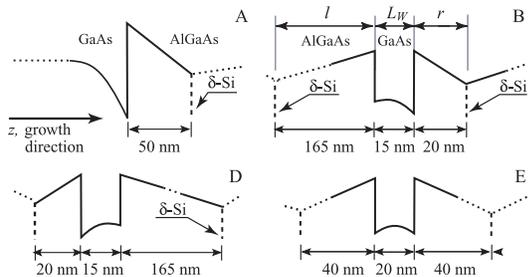}
\caption{Band profile of QWs and doping position.}
\label{figure02}
\end{figure}

The  experiments here are carried out on molecular-beam epitaxy (110)-grown
Si-$\delta$-doped $n$-type  GaAs$/$Al$_{0.3}$Ga$_{0.7}$As
structures. 
The mobility $\mu$ and carrier density $n_s$ measured in the dark at 4.2 K 
are between $0.8\times 10^5$ and $2\times 10^5$~cm$^{2}$/Vs and 
between $10^{11}$~cm$^{-2}$ and $10^{12}$~cm$^{-2}$, respectively. 
The conduction-band profiles of the investigated structures
together with the corresponding $\delta$-doping positions and QW widths $L_W$
are shown in Fig.~\ref{figure02}.
The structures essentially differ in their doping profile:
Sample A is a single heterojunction and has the strongest asymmetry,
stemming from the triangular confinement potential.
In samples B and D, the doping layers are
asymmetrically shifted off the QW center either to the left  or to
the right, respectively. This asymmetric doping
yields an asymmetric potential profile inside the QWs (see Fig.~\ref{figure02}).
All QW structures have 10~QWs. A detailed sample
description can be found in Ref.~[\onlinecite{PRL08}].
Sample E was grown fully symmetric containing Si-$\delta$-sheets, placed in the center of each barrier
between adjacent  QWs. Samples grown along $z\parallel[110]$ were square shaped
with the sample edges  of 5~mm length oriented along
$x\parallel[1{\bar 1}0]$ and $y\parallel[00{\bar 1}]$.
To measure photocurrents, ohmic contacts
were alloyed in the center of each  sample edge.
The MPGE was investigated at room temperature and at low temperatures in an optical
cryostat, which allowed the variation of the temperature in a range
of 4.2~K to 293~K.

\begin{figure}[t]
\includegraphics[width=0.95\linewidth]{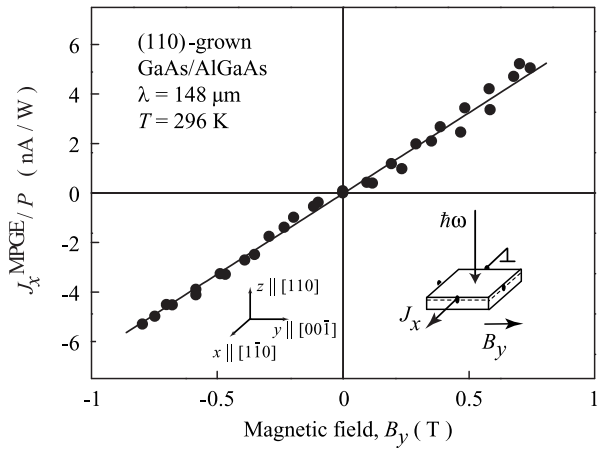}
\caption{Magnetic field dependence of the photocurrent measured in
(110)-grown GaAs/AlGaAs heterojunction
at room temperature with the magnetic field ${\bm B}$
parallel to the $y$ axis. Data are obtained for normally incident
unpolarized radiation of $P \sim 25$~kW at wavelength
$\lambda = 148\:\mu$m.
} \label{figure03}
\end{figure}

An external magnetic field is applied using a conventional 
electromagnet either  in plane, parallel to $y$, or normal to
the QW plane. The field was varied from $B=-0.8$~T to $B=0.8$~T.
The geometry of the experiment is sketched in the inset of Fig.~\ref{figure03}.
The photocurrent is measured in unbiased structures via the voltage drop across
a 50~$\Omega$ load resistor and recorded with a storage oscilloscope. 
In the experiments, the plane of polarization of the radiation, incident on the
sample, was rotated by applying $\lambda/2$ plates, which enabled us to vary
the azimuthal angle $\alpha$ from $0^\circ$ to $180^\circ$ covering all
possible orientations of the electric field vector in the QW plane.
To obtain an unpolarized radiation we used a brass cone of
150~mm length with an angle of 4$^\circ$, to depolarize the radiation due to
multiple reflections in the cone.

%%%%%%%%%%%%%%%%%%%%%%%Figure_02%%%%%%%%%%%%%%%%%%%%%%%%%%%%%%%%%%%%%%%%

The spin dynamics in the samples is investigated by means of time-resolved Kerr rotation (TRKR).
Here, a circularly polarized pump pulse excites spin-polarized electron-hole pairs 
within the QWs. A weaker, time-delayed probe pulse tracks the nonequilibrium spin 
polarization within the sample via the polar magneto-optic Kerr effect: if there is 
a spin polarization normal to the sample plane, the polarization axis of the reflected 
light beam will be tilted by a small angle proportional to the spin polarization. For 
these measurements, the samples are mounted on the cold finger of a He flow cryostat 
with optical access, allowing for sample temperatures between 4~K and room temperature. 
A pulsed Ti-Sapphire laser system generating 600~fs pulses is used for excitation and 
probing. The central wavelength of the laser is tuned above the absorption onset of the QW 
to excite electrons slightly above the Fermi energy. 
For temperature-dependent measurements, this means
that the laser wavelength has to be adjusted in order to follow the temperature-dependent 
absorption onset of the QWs. The laser pulse train is split at a beam splitter, 
and one part of the pulse train is delayed with respect to the other via a mechanical 
delay line. An achromatic quarter-wave plate is used to circularly polarize the pump beam, 
the time-delayed probe beam is linearly polarized. Pump and probe beams are focused 
onto the sample surface at near-normal incidence with an achromat, resulting in a laser 
spot size of about 80~$\mu$m. The Kerr rotation of the reflected probe beam is analyzed 
using an optical bridge detector, and this signal is measured as a function of the 
delay between the pump and probe pulses.  The pump beam is modulated with a flywheel 
chopper, and lock-in detection of the Kerr signal is used to increase the sensitivity.
In order to study the photocarrier dynamics in our samples, time-resolved photoluminescence
(TRPL) measurements were performed. In these measurements, the sample is nonresonantly 
excited well above the Fermi energy by the same Ti-Sapphire laser system used for the TRKR
measurements. The excitation density is 130~W/cm$^2$. The PL emitted from the sample is
collected by an achromat and analyzed by a Hamamatsu streak camera system, synchronized to
the Ti-Sapphire laser. In order to evaluate the time-resolved PL data, the time-resolved 
spectra are averaged over a spectral window of 40~meV, centered around the maximum of 
the PL emission from the QWs. For temperature-dependent measurements, this window is 
accordingly shifted to lower energy as the PL energy decreases.

%%%%%%%%%%%%%%%%%%%%%%%%%%%%%%%%%Figure_03%%%%%%%%%%%%%%%%%%%%%%%%%%%%%%%%%%%%%%

\section{Experimental results}\label{SecIV}
\label{sresults}

Irradiating  samples A, B and D by unpolarized
radiation at normal incidence
we observe a photocurrent 
perpendicular
to the in-plane magnetic field $\bm B$ (transverse geometry).
The  photocurrent pulses duration is about 100~ns,  which
corresponds to the  terahertz laser pulses length. 
Figure~\ref{figure03} shows the magnetic field dependence of
the photocurrent, detected in the GaAs/AlGaAs heterojunction (Sample A).
The photocurrent is proportional to the magnetic field strength
and its sign depends on the magnetic field direction.
The MPGE current has also been detected applying linearly or circularly polarized radiation,
but now also in direction along $\bm B$ (longitudinal geometry).
All these results are in agreement with Eqs.~(\ref{MPGE_Cs})
for  QWs with broken structure inversion.
For sample~E we do not observe any photocurrent as it is expected
for fully symmetric QWs, which do not have structural inversion asymmetry.
Recently, we demonstrated that QWs in the symmetrically doped sample~E
are indeed symmetrical due to the low growth temperature 
used for preparation of (110)-oriented GaAs structures, which 
suppress the segregation process~\cite{PRL08}.

The illumination of samples A, B and D with linearly or circularly polarized
radiation results in a photocurrent even for zero magnetic field. These
currents are due to the linear and circular photogalvanic effects and will be
considered in  Sec.~\ref{scpge}.
As our experiments here are focused on the MPGE, we eliminate the background by
\begin{equation}
\label{MPGE}
J^{\rm MPGE} = [J(\bm{B}) - J(-\bm{B})]/2.
\end{equation}
We note that in the theoretical parts of the paper the current density $\bm{j}^{\rm MPGE}$ is used while
in experiment the electric current $\bm{J}^{\rm MPGE}$ is measured
which is proportional to the current density $\bm{j}^{\rm MPGE}$.

%%%%%%%%%%%%%%%%%%%%%%%%%Figure_04%%%%%%%%%%%%%%%%%%%%%%%%%%%%%%%%%%%%%%%%%%%
\begin{figure}[h]
\includegraphics[width=0.8\linewidth]{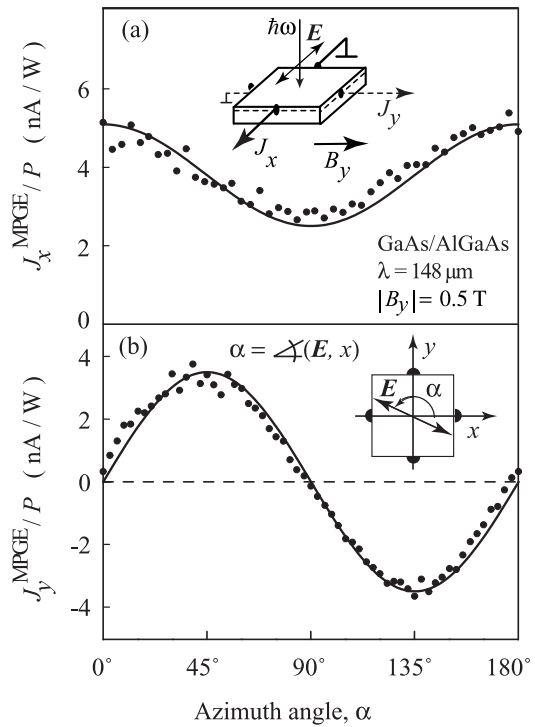}
\caption{ MPGE photocurrent  as a function of angle $\alpha$ measured
along (a) x and (b) y axes
for magnetic field applied in $y$-direction.
Photocurrent is excited by linearly polarized radiation with wavelength $\lambda$~=~148~$\mu$m
and power $P \sim 5$~kW.  
Full lines are fits to Eqs.~\protect(\ref{MPGE_Cs}).
The inset shows the experimental geometry. An additional inset in the lower
panel displays the sample and the radiation 
polarization viewing from the source of radiation side.}
 \label{figure04}
\end{figure}

\begin{figure}[t]
\includegraphics[width=0.85\linewidth]{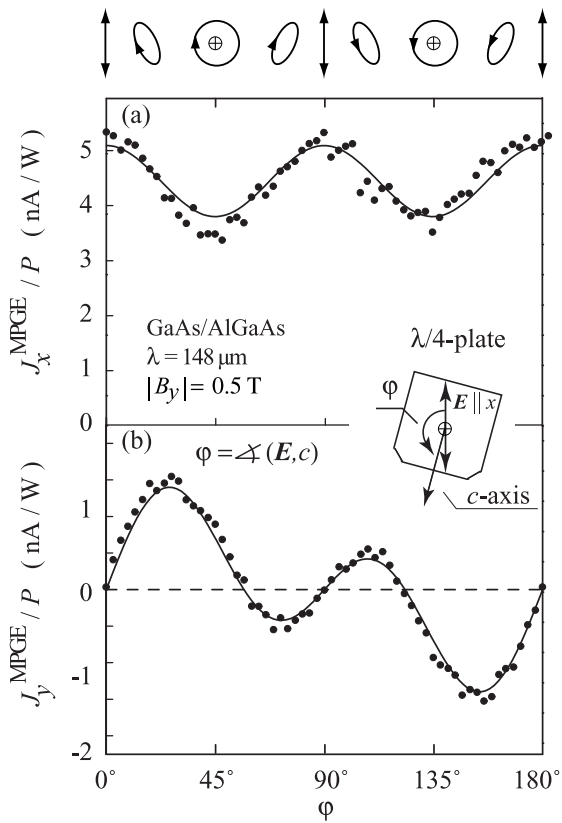}
\caption{
MPGE photocurrent  as a function of angle $\varphi$ measured 
along (a) $x$ and (b) $y$ axes 
for magnetic field applied in $y$ direction.
Photocurrent is excited by elliptically polarized radiation with wavelength $\lambda$~=~148~$\mu$m
and power $P \sim 5$~kW. The ellipticity  of the radiation is varied by passing
linearly polarized laser radiation  through a quarter-wave plate
(see inset). Full lines are fits of the photocurrent
to the sum of Eqs.~(\protect \ref{MPGE_Cs}) and (\protect \ref{MPGE_Cs2})  with corresponding
polarization dependence given by Eqs.~(\protect \ref{plin}) - (\protect \ref{circ}).
On top, the polarization ellipses  corresponding to various phase angles $\varphi$ are
plotted.}
 \label{figure05}
\end{figure}

For polarized radiation, the photocurrent is observed  in both perpendicular
(transverse geometry, Figs.~\ref{figure04}a and~\ref{figure05}a) and parallel 
(longitudinal geometry, Figs.~\ref{figure04}b and~\ref{figure05}b) to
the magnetic field $\bm B$. The results obtained for $\lambda = 90~\mu$m,
$\lambda = 148~\mu$m and $\lambda = 280~\mu$m  are qualitatively the same. Therefore,
we present only  data obtained for  $\lambda = 148~\mu$m.
As discussed above, the contributions are characterized by
different dependencies of the photocurrent magnitude and direction
on the radiation polarization state and the orientation of the
magnetic field with respect to the crystallographic axes. As a
consequence, a proper choice of the experimental geometry allows one to
investigate each contribution separately.

Figure~\ref{figure04} shows the dependence of the photocurrent strength on the
orientation of the polarization plane of linearly polarized radiation, given by
the angle $\alpha$ for both geometries. 
The data are presented for a single heterojunction (sample~A)
which is single-side doped and belongs to the point group C$_{{\rm s}}$. 
From Fig.~\ref{figure04} we can clearly see that $J_x$ is proportional to $\cos\,2\alpha$ and 
$J_y$ is proportional to $\sin\,2\alpha$. 
According to Eqs.~(\ref{MPGE_Cs}) three photocurrent
contributions  proportional to $S_1$, $S_2$ and $S_8$ are allowed in
this configuration. The first contribution in the transverse geometry
is the same as the one detected for unpolarized radiation 
which has the same intensity. Two other contributions are proportional to $I
B_{y} \cos\,2\alpha$ and $I  B_{y}  \sin\,2\alpha$ for the
transverse photocurrent $J_x$ and the longitudinal photocurrent $J_y$,
respectively, in full agreement with the experiments in Fig.~\ref{figure04}.

%%%%%%%%%%%%%%%%%%%%%%%%%%%%%%%%%%Figure_05%%%%%%%%%%%%%%%%%%%%%%%%%%%%%%%%%%%%%%%

Applying elliptically polarized radiation we also observed a magnetic field induced photocurrent.
The dependence of $J^{{\rm \: MPGE}}$ as a function of %radiation helicity 
the angle $\varphi$ is shown in Fig.~\ref{figure05} for sample~A.
The data for transverse geometry shown in Fig.~\ref{figure05}a can be well fitted  by  Eqs.~(\ref{MPGE_Cs}) taken
into account Eqs.~(\ref{plin}) and (\ref{plinprime}). We note that 
the curves are fitted with the same values of $S_1$, $S_2$, and $S_8$ as we used to describe the
experiments with linearly polarized radiation.
In longitudinal geometry, however, {\em elliptically} polarized light yields
an additional helicity dependent current in agreement with
Eqs.~(\ref{MPGE_Cs2}) containing the term proportional to parameter $S_{11}$ and
radiation helicity $P_{circ}$. While the photocurrent contributions described
by $S_2$ and $S_8$  result in a current for linear or elliptical polarization,
the photocurrent described by the coefficient $S_{11}$ vanishes for linear
polarization and assumes its maximum at circular polarization. The polarity of
this photcurrent changes upon reversal of the applied magnetic field as well as
by changing the helicity from right- to left-handed. The polarization
behavior of the current is well described by $j_{y} \propto I B_{y} P_{circ}$.
So far a magnetic-field induced photocurrent proportional to $P_{\rm circ}$ has
been observed for (001)-grown QWs. It is caused by
spin-galvanic effect~\cite{Nature02} generated by the optical
orientation of carriers, subsequent Larmor precession of %the
oriented electronic spins and asymmetric spin relaxation
processes. Though, in general, the spin-galvanic current does not
require the application of a magnetic field, it may  be considered as
a magneto-photogalvanic effect under the above experimental
conditions.

%%%%%%%%%%%%%%%%%%%%%%%%%%%%%Figure_06%%%%%%%%%%%%%%%%%%%%%%%%%%%%%%%%%%%%
\begin{figure}%[t]%[width=4cm]
\includegraphics[width=0.95\linewidth]{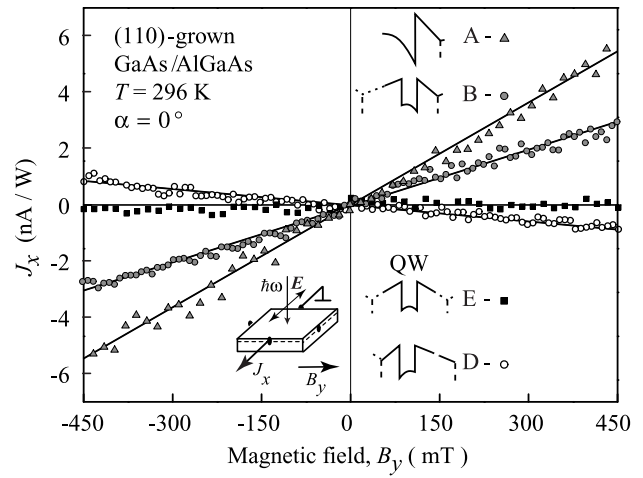}
\caption{Magnetic field dependences of the photocurrents measured in $x$-direction
for the radiation polarized along $x$ and the
in-plane magnetic field $\bm{B} \parallel y$.
The magnetic field independent background (see Sec.~\protect \ref{scpge})
is subtracted. The left inset   shows the experimental geometry. Four right insets show
the band profile and the $\delta$-doping position of the investigated samples.}
\label{figure06}
\end{figure}

The experiment displayed in Fig.~\ref{figure06} (samples A to E) shows that the
magnitude of the $J(B_y)$-slope strongly depends on the  doping profile.
Furthermore, if the doping layers are asymmetrically shifted off the barrier
center from the left  to the right side of QWs (from sample B to D,  see
Fig.~\ref{figure02}), the slope of the photocurrent gets reversed, too (see
Fig.~\ref{figure06}).  The data are presented for room temperature and linearly
polarized radiation with $\bm{e} \parallel x$.  For this geometry, the MPGE
current   $j^{{\rm \: MPGE}}_x /I = (S_1 + S_2) B_y + S_5 B_z $
is phenomenologically determined by the
coupling of the $x$-component of the current polar vector with components of
the axial vector of the magnetic field, because $|e_x|^2$ is an invariant in
(110)-grown structure.  From the point of view of the phenomenological theory
this situation is similar to the case of unpolarized radiation.
Due to the symmetry arguments, presented above, the MPGE current for an in-plane
$\bm{B}$-field is only proportional to the SIA coefficient. To check this, we
rotated the sample by 90 degree so that the $\bm{B}$-field is oriented along
$x$ and the current is measured along the $y$-axis. In this geometry we
detected a signal of the same magnitude and polarization dependence as
before. This proves the axial symmetry of the signal expected for SIA contribution.
Our observations also demonstrate that the position of the doping layer
can be effectively used for tuning  the structure asymmetry strength.
In particular, the absence of
the photocurrent for the in-plane magnetic field in sample~E
indicates that the QW is highly symmetric and
lacks the structure  asymmetry.
The opposite sign of the MPGE observed for samples B and C
having the same QW width
demonstrates that the sign of  $(S_1+S_2)$ 
can be inverted by putting the doping layer
to the other side of the QW.

\begin{figure}[width=4cm]
\includegraphics[width=0.95\linewidth]{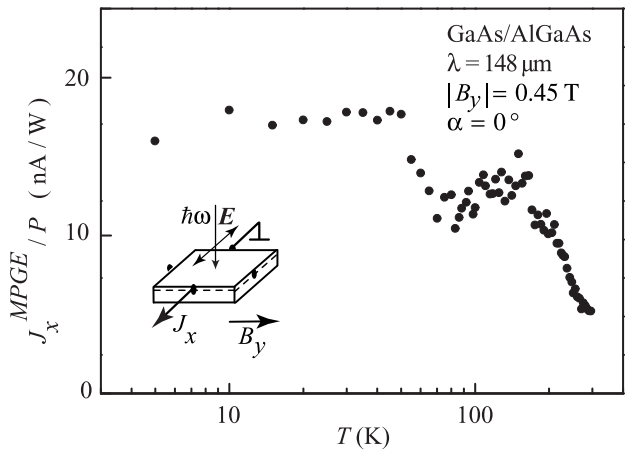}
\caption{ Temperature dependence of the transverse magnetic field induced
photocurrent $J^{{\rm \: MPGE}}_x$. Data are obtained at an 
in-plane magnetic field of $|B_y|= 0.6$~T and
radiation with $P \sim 10$~kW polarized along $x$. }
\label{figure07}
\end{figure}

The experiments discussed above are presented for room temperature. However,
all experimental features, such as magnetic field and polarization dependences,
persist at least down to liquid helium temperature. Figure~\ref{figure07} shows the
temperature dependence for the photocurrent in response to the radiation with
$\lambda=148\mu$m for an in-plane magnetic field $B_y$. The data show that
cooling the sample results in  a significant increase of the photocurrent
strength. We note that a similar temperature dependence for
magnetic-field-induced photocurrents has been previously reported for
(001)-oriented GaAs QWs  and attributed to the conversion of pure spin
currents, generated by THz radiation, into an electric current due to the
equilibrium spin orientation caused by the Zeeman spin
splitting~\cite{naturephysics06}. The observation of a substantial MPGE
response at low temperature on the one hand allows the investigation of the
temperature dependence of SIA
and on the other hand increases the sensitivity of the method. 
Also at low temperatures we did not observe a MPGE signal for an in-plane
magnetic field  in sample E, demonstrating that the structure remains
symmetric. This is an important result in respect to a recent work on
gate-dependent Kerr measurements, where surprisingly a large temperature
dependence of SIA has been reported~\cite{Eldridge07,Eldridge08}.

%%%%%%%%%%%%%%%%%%%%%%%%%%%%Figure_07%%%%%%%%%%%%%%%%%%%%%%%%%%%%%%%%%

\begin{figure}%[t]%[width=4cm]
\includegraphics[width=0.95\linewidth]{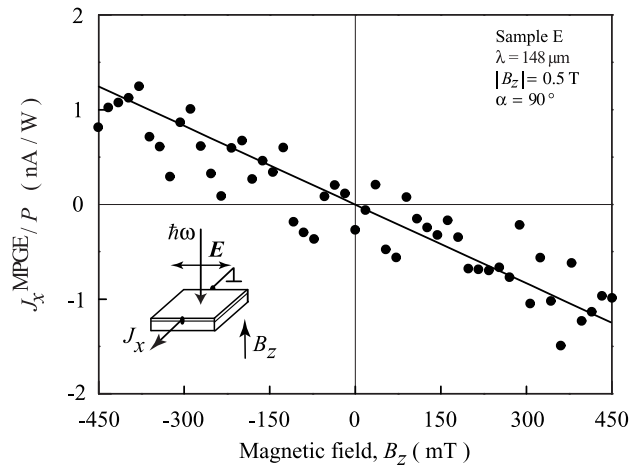}
\caption{Magnetic field dependences of the photocurrent $J_x$
for sample E measured for the radiation polarized along $x$ and
a magnetic field perpendicular to the QWs. The magnetic field independent
background discussed in the last section is subtracted. }
\label{figure08}
\end{figure}

\begin{figure}[h]
\includegraphics[width=0.95\linewidth]{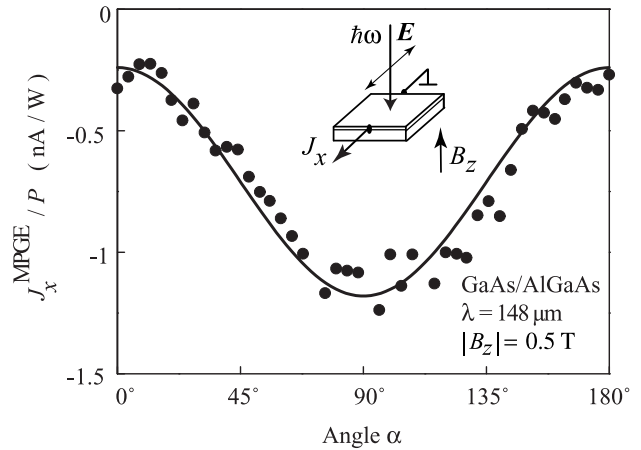}
\caption{ Photocurrent $J^{{\rm \: MPGE}}_x$ as a function of angle $\alpha$ measured at
normal incidence for a magnetic field
perpendicular to the QWs.
Photocurrent is excited by linearly polarized radiation with wavelength $\lambda$~=~148~$\mu$m
and power $P \sim 5$~kW.  Full lines are fits to Eqs.~\protect(\ref{MPGE_Cs}).
The inset shows the experimental geometry.}
 \label{figure09}
\end{figure}

Equations~(\ref{MPGE_Cs}) and~(\ref{MPGE_C2v}) show that for a magnetic field
oriented perpendicular to the quantum well plane, the MPGE is allowed for both
symmetrical and asymmetrical structures. 
The MPGE photocurrent for $B_z$ is indeed observed for all
samples. While no MPGE is observed for in-plane magnetic field  in sample E,
a sizable effect is detected for $\bm{B}$ applied normal to the QW plane, see Fig.~(\ref{figure08}).  The
signals observed for an out-of-plane $B_z$ field  stem from the BIA term [see
Eq.~(\ref{MPGE_C2v})].
Hence, measurement of the MPGE gives us an experimental handle to analyze the inversion
asymmetry in (110)-oriented structures.
Figure~\ref{figure09} shows the polarization dependence of the photocurrent detected
in the single heterojunction, sample~A, being in good agreement with Eq.~(\ref{MPGE_Cs}).

%%%%%%%%%%%%%%%%%%%%%%%%%%%%Figure_08%%%%%%%%%%%%%%%%%%%%%%%%%%%%%%%%%%%%%%

%%%%%%%%%%%%%%%%%%%%%%%%%%%Figure_09%%%%%%%%%%%%%%%%%%%%%%%%%%%%%%%%%%%%%%%

\section{Time-resolved experiments}
\label{stimeresolved}

The structure inversion asymmetry determines  the Rashba spin splitting
and therefore controls the D'yakonov-Perel' (DP)
relaxation~\cite{Dyakonov08,Fabian08} for spins aligned along the $z$ direction. Any
variation of SIA, e.g., due to asymmetric doping,  should result in a variation of
the spin relaxation time. To directly demonstrate this connection, we compare spin 
relaxation rates measured in the symmetrically doped QW, sample E, and the
asymmetrically doped QW, sample B. In the previous section, the potential profiles of samples E 
and B were identified via the MPGE indeed as symmetric and asymmetric, respectively.
We extract the spin lifetime $\tau_s$  from
time-resolved Kerr rotation (TRKR).

\begin{figure}
 \includegraphics[width= 0.4\textwidth]{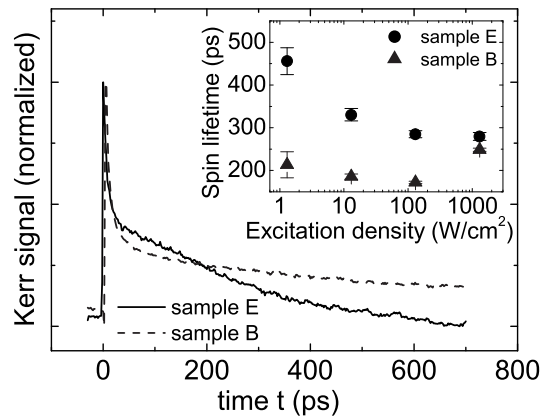}
  \caption{Normalized TRKR traces measured on sample E at 4~K, with high (1300~W/cm$^2$, black line)
  and low (1~W/cm$^2$, red dotted line) excitation density.
  The inset shows the spin lifetime as a function of the excitation 
  density at 4~K for samples E(black dots) and B(red triangles).}
\label{figure10}
\end{figure}

We first discuss the spin lifetime in both, the symmetrically-grown sample E and the
asymmetric sample B, as a function of
the excitation density. Figure~\ref{figure10} shows two normalized TRKR traces
measured on sample E at 4~K, using high and low excitation densities. At zero time delay, the pump
pulse creates spin-polarized photocarriers, resulting in a maximum Kerr signal. This signal
partially decays very rapidly within the first few picoseconds, then the decay becomes much slower. 
We attribute this first, rapid decay to the spin relaxation of the photogenerated holes, 
which typically lose their spin orientation within a few picoseconds in QWs~\cite{Damen91}. 
We attribute the slower decay to the spin relaxation and recombination of the photogenerated electrons. 
It is clearly visible from the traces that the Kerr signal, and with it the electron 
spin polarization, decays more rapidly at higher excitation density.

The inset of Fig.~\ref{figure10} shows this increase of the spin lifetime for
a reduction of the excitation density in sample E for a wide range of data. 
The spin lifetime in the asymmetric sample B, however, shows a different 
dependence on the excitation density: it is largest at the highest excitation density 
used in the measurements, then first decreases by about 30~percent as 
the excitation density is reduced. At lower excitation density, a slight 
increase is observed.
Two factors may contribute to the marked increase of the spin lifetime with 
reduced excitation density in sample E:

First, the measured Kerr signal is proportional to
the spin polarization within the sample. In  undoped samples, a spin polarization
may only persist during the lifetime of the generated photocarriers. Therefore,
in samples where electron spin relaxation is slow compared to photocarrier recombination,
the decay of the Kerr signal will reflect the lifetime of the photocarriers. In n-doped samples, 
like our structures, however, an electron spin polarization may remain after photocarrier recombination, 
as  photogenerated holes partially recombine with unpolarized, resident electrons. The Kerr signal 
will therefore reflect a combination of electron spin relaxation and photocarrier recombination. 
If the excitation density is very low compared to the doping concentration within the sample, 
electron spin relaxation will dominate the decay of the Kerr signal, for high excitation density, 
the photocarrier recombination will increase the observed decay rate of the Kerr signal.

Second, the DP mechanism, which typically dominates the electron spin relaxation
in GaAs QWs, is suppressed for spins aligned along the growth direction in the 
symmetrically-grown sample E. In its absence, other spin relaxation mechanisms become relevant. 
For low sample temperatures, we need to consider the  Bir-Aronov-Pikus (BAP) mechanism. It describes 
the spin relaxation of electrons via their interaction with (unpolarized) holes. The strength of 
the BAP mechanism scales with the density of the photocreated holes, a reduction of the excitation 
density will therefore decrease
its influence. Recent spin noise spectroscopy measurements~\cite{Oe08} demonstrate that the BAP 
mechanism  strongly reduces the observed  spin lifetime in (110)-grown QW systems, 
which may reach values above 20~ns in the absence of photogenerated holes.

From the excitation-density-dependent data, we may conclude that in sample B, the spin 
relaxation is dominated by the DP mechanism, as the asymmetric growth leads to SIA and a 
corresponding Rashba spin splitting. The BAP mechanism is, by comparison, far less important 
for spin relaxation in this sample, therefore the reduction of the photogenerated hole density 
does not significantly increase the spin lifetime.  The increase of the spin lifetime for the 
highest excitation density may  be due to the increase in initial spin polarization, 
as observed previously in (001)-grown 2D electron systems~\cite{Stich}.

%%%%%%%%%%%%%%%%%%%%%%%%%%%%%%Figure_10%%%%%%%%%%%%%%%%%%%%%%%%%%%%%%%%%%%%%%%%%%%%%%%

%%%%%%%%%%%%%%%%%%%%%%%%%%%%%%Figure_11%%%%%%%%%%%%%%%%%%%%%%%%%%%%%%%%%%%%%%%%%%%%%%
\begin{figure}
 \includegraphics[width= 0.4\textwidth]{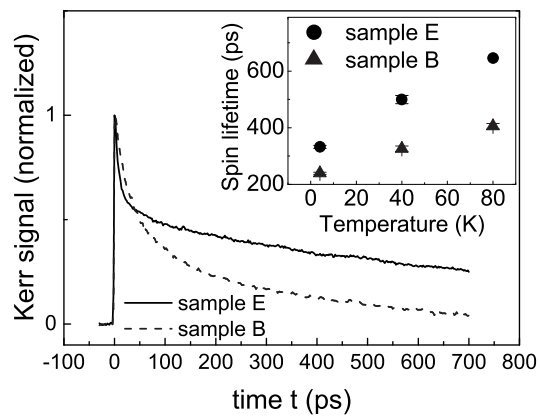}
  \caption{Normalized TRKR traces measured on samples E (black line) and
  B (red dotted line) at 40~K with high excitation density.  The inset shows the spin
  lifetime as a function of the sample temperature for samples E (black dots) and B
  (red triangles).}
\label{figure11}
\end{figure}

\begin{figure}
 \includegraphics[width= 0.4\textwidth]{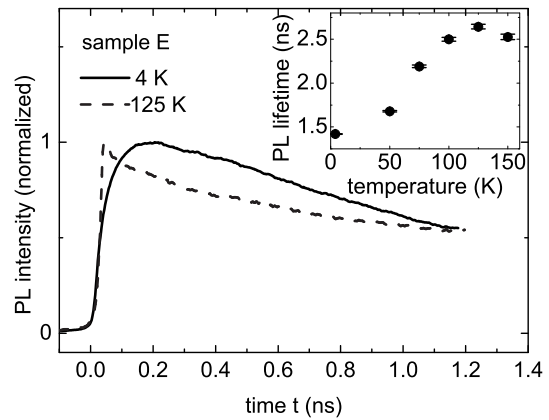}
  \caption{Spectrally integrated TRPL traces measured on sample E for
  4~K (black line) and 125~K (red dotted line). The inset shows the
  PL lifetime as a function of temperature for sample E.}
\label{figure12}
\end{figure}

Next, we investigate the spin lifetime in sample B and E as a function of the sample temperature.
Figure~\ref{figure11} shows
two normalized TRKR traces measured on these samples at 40~K, using a high excitation density. 
Both samples show a partial decay of the Kerr signal during the first few ps after excitation, 
which we again attribute to the spin relaxation of photoexcited holes. After this initial decay, 
the Kerr signal of sample~E decays significantly more slowly than that of sample~B. The inset in 
Fig.~\ref{figure11} shows the spin lifetimes of the two samples as a function of temperature. These 
spin lifetimes were determined by an exponential fit to the TRKR traces starting at t=200~ps in order 
to exclude the hole spin relaxation. We note that the symmetrically-grown sample E shows a spin 
lifetime which is about 50~percent larger than that of sample B, within the whole temperature range 
investigated here. This observation clearly confirms our interpretation that in the the symmetrically-grown 
sample E, the Rashba spin-orbit field is absent, and spin relaxation via the DP mechanism is suppressed, 
while it is still present, and dominates the spin relaxation, in the asymmetric sample B. All the 
temperature-dependent measurements were performed at rather high excitation density (130~W/cm$^2$), 
therefore, the measured spin lifetime for sample E is significantly lower than the intrinsic limit 
in the absence of photocarriers. As the inset  in Fig.~\ref{figure11} shows, the spin lifetime in 
both samples increases monotonically with increasing temperature. We can identify two factors, 
which contribute to this increase:

(i) At higher excitation density, the TRKR decay is increased by photocarrier recombination. In 
order to study the temperature dependence of the photocarrier lifetime, we performed TRPL 
measurements on sample E as a function of temperature.
Figure~\ref{figure12} shows normalized, spectrally averaged TRPL traces measured on sample E at 
4~K and 125~K. After pulsed excitation at t=0~ns, the low-temperature TRPL trace shows an increase 
of the PL intensity within the first 150~ps, then the PL intensity decreases monotonically. This 
behavior is  typical for nonresonant excitation of a QW at low temperatures~\cite{Dyakonov08}, as 
the photoexcited electrons and holes first have to reduce their momenta via scattering to enter 
the so-called light cone~\cite{Shields} before they may recombine radiatively. The trace taken at 
125~K does not show this initial increase, but monotonically decreases after excitation. At these 
elevated temperatures, carrier-phonon scattering is much more pronounced, allowing for faster 
scattering of the photoexcited carriers into and out of the light cone. The latter effect leads 
to an increase of the PL lifetime as the temperature is increased.%Schüller
The photocarrier lifetime is extracted from the TRPL data by  an exponential fit to the TRPL traces 
starting at t=0.5~ns in order to exclude the influence of initial thermalization and carrier-phonon 
scattering. The inset in Fig.~\ref{figure12} shows this photocarrier lifetime as a function of temperature. 
 It increases by about 90~percent as the sample temperature is raised from 4~K to 125~K, then decreases 
 again slightly. In the temperature range where we have studied the spin lifetime in our samples, the 
 photocarrier lifetime increases monotonically, therefore reducing the TRKR decay via photocarrier recombination.
We note that the photocarrier lifetime after near-resonant excitation is typically shorter
than that observed after nonresonant excitation. Therefore, the TRPL data  give only a
qualitative indication of the photocarrier lifetime under conditions of near-resonant excitation.

(ii) The BAP mechanism, which limits the spin lifetime, especially in the symmetrically-grown sample~E, 
may become less important in our structures as the sample temperature is increased. The temperature 
dependence of the BAP mechanism in QWs was recently investigated theoretically by Zhou et al.~\cite{WuBAP} 
using the kinetic spin-Bloch equations approach. In their model, they observe an increase of the 
BAP mechanism as the sample temperature is raised, which is in contrast to our observations. One possible
explanation for this contradiction may be the experimental conditions: in the TRKR measurements, 
the holes are created locally by the pump beam. With increasing temperature, the spatial overlap 
between electrons and holes is reduced~\cite{Dohrmann}, leading to the observed increase in 
photocarrier lifetime and spin lifetime.

%%%%%%%%%%%%%%%%%%%%%%%%%%%%%%Figure_12%%%%%%%%%%%%%%%%%%%%%%%%%%%%%%%%%%%%%%%%%%%%%%

\section{Photocurrents at zero magnetic field}
\label{scpge}

\begin{figure}[h]
\includegraphics[width=0.9\linewidth]{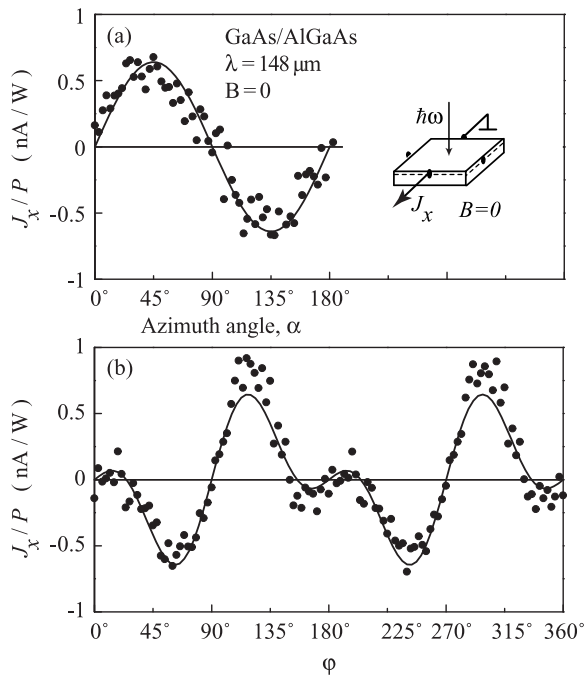}
\caption{ Photocurrent, measured at zero magnetic field along the crystallographic
direction $x \parallel [1\bar{1}0]$ as a function of (a) azimuth angle $\alpha$ and (b) 
the angle $\varphi$.
The photocurrent is excited by radiation with wavelength $\lambda$~=~148~$\mu$m
and power $P \sim 5$~kW.
Full lines are fits to Eqs.~\protect(\ref{CPGELPGE}).
The inset shows the experimental geometry.}
 \label{figure13}
\end{figure}

Investigating the MPGE one should take into account
that in (110)-oriented structures  optical excitation may generate
other photocurrents at normal incidence, even for zero magnetic field.
The MPGE, however, can  easily be extracted from the total photocurrent by treating
the data after Eq.~(\ref{MPGE}). Indeed, only the MPGE being odd in $B$,
changes its direction when the magnetic field direction is inverted.
In the infrared/terahertz spectral range  there can be two  sources of
photocurrents at homogeneous excitation  which occur simultaneously and may be
of the same order of magnitude as the MPGE.
These are the linear and circular photogalvanic effects and
the photon drag effect~\cite{Ivchenkobook2,GanichevPrettl}. 
Our experiments demonstrated that in the investigated samples the photocurrent 
at zero magnetic field is
in most cases caused by the linear photogalvanic effect.
The photocurrent is observed in both, $x$ and $y$ directions for linearly as well as for
elliptically polarized radiation.
Figure~\ref{figure13} shows the photocurrent detected in sample A for room temperature and
zero magnetic field as a function of the azimuth angle of linear polarization (Fig.~\ref{figure13}a)
and radiation helicity~(Fig.~\ref{figure13}b).
The latter plot reveales that in $x$ direction, being perpendicular to the mirror reflection plane
$m_1$, the  circular photogalvanic effect  overweight the linear photogalvanic effect.
The data are in a good agreement with  Eqs.~(\ref{CPGELPGE}).

%%%%%%%%%%%%%%%%%%%%%%%%%%%%%%%%%%%Figure_13%%%%%%%%%%%%%%%%%%%%%%%%%%%%%%%%%%%

\section{Summary}

We have studied  photocurrents  in $n$-doped zinc-blende based
(110)-grown QWs generated by Drude absorption of normally
incident terahertz radiation in the presence of an in-plane and out of plane
magnetic field. The results agree with the phenomenological
description based on the symmetry. The observation of polarization-dependent
as well as polarization-independent photocurrents reveal that
 both, an asymmetry of photoexcitation and asymmetry of energy relaxation
contribute substantially to the MPGE in (110)-grown QWs. We show that the MPGE
provides a tool to probe the degree of the structural inversion asymmetry which
defines the spin relaxation in (110)-grown quantum wells.
Parallel to the MPGE experiments, we also investigated spin relaxation
applying  time-resolved Kerr rotation and luminescence.
As an important result of all our measurements we obtained a zero current
response in the in-plane magnetic field
and the longest spin relaxation time from the 
symmetrically doped QWs, which set an upper limit of
spin dephasing in GaAs QWs.
This is in contrast to (001)-grown structures, where such a  growth
procedure  results in a
substantial SIA~\cite{prBRD,APL09}.
This essential difference stems most likely from increased segregation 
at high growth termperatures, and,
subsequently, the diffusion length of dopant atoms.
Indeed, the growth temperature of high-quality (001)-oriented QWs is higher
than 600$^\circ$C, while (110)-structures are grown at 480$^\circ$C~\cite{W2}.
High growth temperature of (001)-oriented heterostructures leads to substantial
dopant migration and results in structure asymmetry of nominally symmetrically
doped QWs. The investigation of the MPGE, in particular the sign inversion by
reversing of structural asymmetry and the zero current response in MPGE of
symmetrical  structures, provides an effective access to  study  the symmetry
of (110)-oriented QWs. The observed rise of the photocurrent strength with decreasing 
temperature demonstrates that the MPGE can be applied  to
investigate the inversion asymmetry in a wide range of temperatures including
technologically important room temperature, where many methods, like weak
localization,  or polarized luminescence, can not be used.
In summary our photocurrent measurements provide the necessary feedback to reliably 
grow structures with long spin relaxation times.

We thank E.L.~Ivchenko and M.W.~Wu for fruitful discussion, as well as V. Lechner and S. Stachel.
The financial support by the DFG and RFBR is gratefully acknowledge.


\begin{thebibliography}{99}
\bibitem{footnote1} Note that C$_{2v}$ point group symmetry
also describes the (001)-grown asymmetric structures. In this case, however,
both mirror reflection planes are perpendicular to the interface plane.
\end{thebibliography}
\end{document}